\documentclass{llncs}
\usepackage{subscript} 

\usepackage{makeidx}  
\usepackage{hyperref} 

\usepackage{algorithmic}
\usepackage{float}
\usepackage[font=small,skip=0pt]{caption}
\usepackage[font=footnotesize]{subcaption}

\usepackage{times}
\usepackage{graphicx,verbatim,color,epsfig,amsmath}
\usepackage[ruled,linesnumbered,vlined]{algorithm2e}
\usepackage{ifthen,array}
\usepackage{verbatim,color,latexsym,amssymb,amsmath,multirow,setspace}
\usepackage{xspace}
\usepackage{setspace}
\usepackage{textcomp}
\usepackage{stmaryrd}

\let\llncssubparagraph\subparagraph
\let\subparagraph\paragraph
\usepackage{titlesec}
\let\subparagraph\llncssubparagraph

\usepackage{hyperref}
\usepackage{enumitem}
\usepackage{caption} 

\usepackage[table]{xcolor}
\definecolor{lightgray}{gray}{0.925}

\usepackage{tikz}

\newcommand{\tn}{\textnormal}
\newcommand{\fml}[1]{{\mathcal{#1}}}

\newcommand{\elplus}{$\fml{EL}^{+}$\xspace}
\newcommand{\elplain}{$\fml{EL}$\xspace}
\newcommand{\hgmus}{\textsc{HgMUS}\xspace}
\newcommand{\emus}{\textsc{eMUS}\xspace}
\newcommand{\elsat}{EL$^{\tn{+}}$SAT\xspace}
\newcommand{\just}{\textsc{Just}\xspace}

\newcommand{\set}[1]{\{ #1 \}}
\DeclareMathOperator*{\nentails}{\nvDash}
\DeclareMathOperator*{\entails}{\vDash}

\usetikzlibrary{shapes,arrows}

\hypersetup{%
    bookmarks=false,    
    pdftitle={},        
    pdfauthor={Joao Marques-Silva},                    
    pdfsubject={TeX and LaTeX},                        
    pdfkeywords={TeX, LaTeX, graphics, images}, 
    colorlinks=true,       
    linkcolor=blue,        
    citecolor=blue,        
    filecolor=black,       
    urlcolor=blue,         
    linktoc=page           
}

\AtBeginDocument{%

\renewcommand*{\algorithmautorefname}{Algorithm}

}

\begin{document}

\title{Efficient MUS Enumeration of Horn Formulae\\[2pt]
  with Applications to Axiom Pinpointing}
\titlerunning{SAT-Based Axiom Pinpointing of $\fml{EL}^+$ Ontologies}  
%
\author{M.\ Fareed Arif\inst{1} \and Carlos Menc\'{\i}a\inst{1} \and
  Joao Marques-Silva\inst{1,2}}
\authorrunning{Arif \& Menc\'{\i}a, \& Marques-Silva}
%
\institute{
CASL, University College Dublin, Ireland\\
\email{muhammad.arif.1@ucdconnect.ie,carlos.mencia@ucd.ie,jpms@ucd.ie}
\and
INESC-ID, IST, ULisboa, Portugal\\
}

\maketitle              

\setcounter{footnote}{0}

\begin{abstract}
The enumeration of minimal unsatisfiable subsets (MUSes) finds a
growing number of practical applications, that includes a wide range
of diagnosis problems. 
As a concrete example, the problem of axiom pinpointing in the
\elplain family of description logics (DLs) can be modeled as the
enumeration of the group-MUSes of Horn formulae. In turn, axiom pinpointing
for the \elplain family of DLs finds important applications, 
such as debugging medical ontologies, of which SNOMED~CT is the best
known example.
The main contribution of this paper is to develop an efficient
group-MUS enumerator for Horn formulae, \hgmus, that finds immediate
application in axiom pinpointing for the \elplain family of DLs.
In the process of developing \hgmus, the paper also identifies
performance bottlenecks of existing solutions.
The new algorithm is shown to outperform all alternative approaches
when the problem domain targeted by group-MUS enumeration of Horn
formulae is axiom pinpointing for the \elplain family of DLs, with a
representative suite of examples taken from different medical
ontologies. 
\end{abstract}

\section{Introduction}

Description Logics (DLs) are well-known knowledge representation
formalisms~\cite{baader-hdbk08}. DLs find a wide range of applications
in computer science, including the semantic web and representation
of ontologies, but also in medical bioinformatics. 
Axiom pinpointing represents the problem of computing one minimal axiom
set (denoted {\em MinA}), which explains a subsumption relation in an
ontology~\cite{schlobach-ijcai03}. 
Example applications of axiom pinpointing include context-based
reasoning, error-tolerant reasoning~\cite{ludwig-jelia14}, and
ontology debugging and revision~\cite{schlobach-jar07,kalyanpur-eswc06}.
Axiom pinpointing for different description logics (DLs) has been
studied extensively for more than a decade, with related work in the
mid
90s~\cite{baader-jar95,schlobach-ijcai03,parsia-www05,meyer-aaai06,baader-ijcar06,baader-ki07,parsia-iswc07,parsia-jws07,schlobach-jar07,baader-krmed08,sebastiani-cade09,baader-jlc10,meyer-rr11,nguyen-dl12,ludwig-ore14,sebastiani-tr15,ams-corr15,mp-tr15}.

The \elplain family of DLs is well-known for being tractable. Despite
being inexpressive, the \elplain family of DLs, concretely by using
the more expressive \elplus, has been used for representing ontologies
in the medical sciences, including the well-known SNOMED~CT
ontology~\cite{spackman-amia97}.
Work on axiom pinpointing for the \elplain family of DLs can be traced
to 2006, namely the CEL tool~\cite{baader-ijcar06}.
Later, in 2009, the use of SAT was proposed for axiom pinpointing in
the \elplain family of
DLs~\cite{sebastiani-cade09,vescovi-phd11,sebastiani-tr15}, concretely
for the more expressive description logic \elplus. This seminal work
proposed a propositional Horn encoding that can be exponentially
smaller than earlier
work~\cite{baader-ijcar06,baader-ki07,baader-krmed08}. Moreover, the
use of SAT for axiom pinpointing for the \elplain family of DLs, named 
\elsat~\cite{sebastiani-cade09,vescovi-phd11,sebastiani-tr15}, was
shown to consistently outperform earlier work, concretely
CEL~\cite{baader-ijcar06}. 
Recent work~\cite{ams-corr15} proposes the EL2MCS tool that builds on these propositional encodings,
but exploits the relationship between axiom pinpointing and MUS
enumeration; concretely, it relies on explicit hitting set dualization~\cite{liffiton-jar08}. 
This tool is evaluated in~\cite{ams-corr15}, where it is shown to achieve conclusive 
performance gains over earlier work.
The relationship between axiom pinpointing and MUS enumeration was
also studied elsewhere~\cite{mp-tr15}. Instead of exploiting hitting
set dualization, this alternative approach exploits the enumeration of
prime implicants~\cite{mp-tr15}.

The main contribution of this paper is to develop an efficient group-MUS 
enumerator for Horn formulae, referred to as \hgmus, that finds
immediate application in axiom pinpointing for the \elplain family of
DLs.
In the process of developing \hgmus, the paper also identifies
performance bottlenecks of existing solutions, in particular
\elsat~\cite{sebastiani-cade09,sebastiani-tr15}.
The new group-MUS enumerator for Horn formulae builds on the large
body of recent work on problem solving with SAT oracles. This
includes, among others, MUS extraction~\cite{blms-aicomm12}, MCS
extraction and enumeration~\cite{mshjpb-ijcai13}, and partial MUS 
enumeration~\cite{pms-aaai13,liffiton-cpaior13,lpmms-cj15}. 
\hgmus also exploits earlier work on solving Horn propositional
formulae~\cite{gallier-jlp84,minoux-ipl88}, and develops novel
algorithms for MUS extraction in propositional Horn formulae.
The experimental results, using well-known problem instances,
demonstrate conclusive performance improvements over all other
existing approaches, in most cases by several orders of magnitude.

The paper is organized as follows. \autoref{sec:prelim} introduces the
notation and definitions used throughout the paper.
\autoref{sec:hmusenum} reviews recent work on MUS enumeration, which
serves as the basis for \hgmus. Afterwards, the new group-MUS
enumerator \hgmus is described in~\autoref{sec:hegmus}.
\autoref{sec:relw} compares \hgmus with existing alternatives.
Experimental results on well-known problem instances from axiom
pinpointing for the \elplain family of DLs are analyzed
in~\autoref{sec:res}. The paper concludes in~\autoref{sec:conc}.

\section{Preliminaries} \label{sec:prelim}

We assume familiarity with propositional logic \cite{sat-handbook09} and consider propositional Boolean formulae in Conjunctive Normal Form (CNF). 
A CNF formula $\fml{F}$ is defined over a set of Boolean variables $V(\fml{F}) = \set{x_1,...,x_n}$ as a conjunction of clauses ($c_1 \land ... \land c_m$). A clause $c$ is a disjunction of literals ($l_{1} \lor ... \lor l_{k}$) and a literal $l$ is either a variable $x$ or its negation $\lnot x$. 
We refer to the set of literals appearing 
in $\fml{F}$ as $L(\fml{F})$. Formulae can be alternatively represented as sets of clauses, and clauses as sets of literals. 
 
A truth assignment, or interpretation, is a mapping $\mu : X
\rightarrow \set{0,1}$, with $X=V(\fml{F})$ also used to represent the
variables of $\fml{F}$.
If all the variables in $X$ are assigned a truth value, $\mu$ is referred to as a \textit{complete} assignment. Interpretations can be also seen as conjunctions or sets of literals. Truth valuations are lifted to clauses and formulae as follows: $\mu$ satisfies a clause $c$ if it contains at least one of its literals. 
Given a formula $\fml{F}$, $\mu$ satisfies $\fml{F}$ (written $\mu \entails \fml{F}$) if it satisfies all its clauses, being $\mu$ referred to as a \textit{model} of $\fml{F}$.

Given two formulae $\fml{F}$ and $\fml{G}$, $\fml{F}$ entails $\fml{G}$ (written $\fml{F} \entails \fml{G}$) iff all the models of $\fml{F}$ are also models of $\fml{G}$. $\fml{F}$ and $\fml{G}$ are equivalent (written $\fml{F} \equiv \fml{G}$) iff $\fml{F} \entails \fml{G}$ and $\fml{G} \entails \fml{F}$. 

A formula $\fml{F}$ is satisfiable ($\fml{F} \nentails \bot$) if there exists a model for it. Otherwise it is unsatisfiable ($\fml{F} \entails \bot$). SAT is the decision problem of determining the satisfiability of a propositional formula. This problem is in general NP-\textit{complete} \cite{cook-stoc71}.

Some applications require computing certain types of models. In this paper, we will make use of maximal models, i.e. models such that a set-wise maximal subset of the variables are assigned value 1:

\begin{definition}\label{mxm}
\textbf{(MxM). } Let $\fml{F}$ be a satisfiable propositional formula, $\mu \entails \fml{F}$ a model of $\fml{F}$ and $P \subseteq X$ the set of variables appearing in $\mu$ with positive polarity. $\mu$ is a maximal model (MxM) of $\fml{F}$ iff $\fml{F} \cup P \nentails \bot$ and for all $v \in X \setminus P$, $\fml{F} \cup P \cup \{v\} \entails \bot$. 
\end{definition}

Herein, we will denote a maximal model by $P$, i.e. the set of its positive literals.

Horn formulae constitute an important subclass of propositional logic. These are composed of Horn clauses, which have at most one positive literal. Satisfiability of Horn formulae is decidable in polynomial time~\cite{gallier-jlp84,itai-jlp87,minoux-ipl88}.

Given an unsatisfiable formula $\fml{F}$, the following subsets represent different notions regarding (set-wise) minimal unsatisfiability and maximal satisfiability \cite{liffiton-jar08,mshjpb-ijcai13}:

\begin{definition} \label{def:mus}
\textbf{(MUS).} $ \fml{M}\subseteq\fml{F}$ is a {\em Minimally Unsatisfiable Subset}
(MUS) of $\fml{F}$ iff $\fml{M}$ is unsatisfiable and
$\forall{c\in\fml{M}}, \fml{M}\setminus\{ c \}$ is satisfiable.
\end{definition}
\begin{definition} \label{def:mcs}
\textbf{(MCS).} $\fml{C}\subseteq\fml{F}$ is a {\em Minimal Correction Subset} (MCS)
iff $\fml{F} \setminus \fml{C}$ is satisfiable and
$\forall{c\in\fml{C}}, \fml{F} \setminus (\fml{C}\setminus\{ c \})$
is unsatisfiable.
\end{definition}
\begin{definition} \label{def:mss}
\textbf{(MSS).} $\fml{S}\subseteq\fml{F}$ is a {\em Maximal Satisfiable Subset} (MSS)
iff $\fml{S}$ is satisfiable and
$\forall{c\in\fml{F}\setminus\fml{S}}, \fml{S}\cup\{c\}$ is
unsatisfiable.
\end{definition}

An MSS is the complement of an MCS. MUSes and MCSes are closely related by the well-known hitting set duality \cite{reiter-aij87,stuckey-padl05,lozinskii-jetai03,slaney-ecai14}: Every MCS (MUS) is an irreducible hitting set of all MUSes (MCSes) of  $\fml{F}$. In the worst case, there can be an exponential number of MUSes and MCSes \cite{liffiton-jar08,osullivan-aaai07}. Besides, MCSes are related to the MaxSAT problem, which consists in finding an assignment satisfying as many clauses as possible. The smallest MCS (largest MSS) represents an optimal solution to MaxSAT.

Motivated by several applications, MUSes and related concepts have
been extended to CNF formulae where clauses are partitioned into
disjoint sets called groups~\cite{liffiton-jar08}.

\begin{definition} \label{def:gmus}
\textbf{(Group-Oriented MUS).} Given an explicitly partitioned unsatisfiable CNF formula $\fml{F} = \fml{G}_0 \cup ... \cup \fml{G}_k$, a group-oriented MUS (or group-MUS) of $\fml{F}$ is a set of groups $ \fml{G} \subseteq \set{ \fml{G}_1, ... , \fml{G}_k }$, such that $\fml{G}_0 \cup \fml{G}$ is unsatisfiable, and for every $\fml{G}_i \in \fml{G}$, $\fml{G}_0 \cup (\fml{G} \setminus \fml{G}_i)$ is satisfiable.
\end{definition}

Note the special role $\fml{G}_0$ (\textit{group-0}); this group consists of \textit{background} clauses that are included in every group-MUS. Because of $\fml{G}_0$ a group-MUS, as opposed to MUS, can be empty. Nevertheless, in this paper we assume that $\fml{G}_0$ is satisfiable.

Equivalently, the related concepts of group-MCS and group-MSS can be
defined in the same way. We omit these definitions here due to lack of
space. In the case of MaxSAT, the use of groups is investigated in
detail in~\cite{hmms-aicomm15}.

\section{MUS Enumeration in Horn Formulae} \label{sec:hmusenum}

Enumeration of MUSes has been the subject of research that can be
traced to the seminal work of Reiter~\cite{reiter-aij87}.
A well-known family of algorithms uses (explicit) minimal hitting set
dualization~\cite{lozinskii-jetai03,stuckey-padl05,liffiton-jar08}. The
organization of these algorithms can be summarized as follows. First
compute all the MCSes of a CNF formula. Second, MUSes are obtained by
computing the minimal hitting sets of the set of MCSes.
The main drawback of explicit minimal hitting set dualization is that,
if the number of MCSes is exponentially large, these approaches will
be unable to compute MUSes, even if the total number of MUSes is
small.
As a result, recent work considered what can be described as implicit
minimal hitting set
dualization~\cite{liffiton-cpaior13,pms-aaai13,lpmms-cj15}. In these
approaches, either an MUS or an MCS is computed at each step of the
algorithm, with the guarantee that one or more MUSes will be computed
at the outset. In some settings, implicit minimal hitting set
dualization is the only solution for finding some MUSes of a CNF
formula.
As pointed out in this recent work, implicit minimal hitting set
dualization aims to complement, but not replace, the explicit
dualization alternative, and in some settings where enumeration of
MCSes is feasible, explicit minimal hitting set dualization may be the
preferred option~\cite{pms-aaai13,lpmms-cj15}.

\begin{algorithm}[t]
{\small 
\DontPrintSemicolon
\SetAlgoNoLine
\LinesNumbered
\SetFillComment
\SetKw{KwNot}{not\xspace}
\SetKw{KwAnd}{and\xspace}
\SetKw{KwOr}{or\xspace}
\SetKwData{false}{{\small false}}
\SetKwData{true}{{\small true}}
\SetKwData{st}{\small{\sl st}}
\SetKwFunction{SAT}{{\normalsize SAT}}
\SetKwFunction{ComputeMxM}{MaximalModel}
\SetKwFunction{ComputeMUS}{ComputeMUS}
\SetKwFunction{ReportMUS}{ReportMUS}
\KwIn{$\fml{F}$ a CNF formula}
\smallskip
\SetAlgoVlined
    $I \gets \{ p_i ~|~ c_i \in \fml{F}\}$ 
    \tcp*[r]{Variable $p_i$ picks clause $c_i$} 

    $\fml{Q} \gets \emptyset$\;
    \While{\true} {

      $(st, P) \gets \ComputeMxM(\fml{Q})$ \;

      \lIf{\KwNot $st$}{ \textbf{return} }
  	
      $\fml{F}' \gets \{ c_i ~|~ p_i \in P\}$ 
      \tcp*[r]{Pick selected clauses} 
  	
  	 \eIf{\KwNot $\SAT(\fml{F}')$}{  
  	   $\fml{M} \gets \ComputeMUS(\fml{F}')$ \;
  	   $\ReportMUS(\fml{M})$\;
  	   $b \gets \{ \lnot p_i ~|~ c_i \in \fml{M} \}$
           \tcp*[r]{Negative clause blocking the MUS}
  	}{
  		$b \gets \{p_i ~|~ p_i \in I\setminus P \}$
           \tcp*[r]{Positive clause blocking the MCS}		
  	}
  	$\fml{Q} \gets \fml{Q} \cup \{b\}$\;
  }
}
\caption{\emus~\cite{pms-aaai13}~/~MARCO~\cite{lpmms-cj15} \label{alg:emus}}
\end{algorithm}

Algorithm \ref{alg:emus} shows the \emus enumeration
algorithm~\cite{pms-aaai13}, also used in the most recent version of
MARCO~\cite{lpmms-cj15}.
It relies on a two-solver approach
aimed at enumerating the MUSes/MCSes of an unsatisfiable formula $\fml{F}$.
On the one hand, a formula $\fml{Q}$ is used to enumerate subsets of $\fml{F}$. This formula 
is defined over a set of variables $I = \{p_i ~|~ c_i \in \fml{F}\}$, each one of them associated with one clause
$c_i \in \fml{F}$. Iteratively until $\fml{Q}$ becomes unsatisfiable, \textsc{eMUS} computes a maximal model $P$ of $\fml{Q}$ and 
tests the satisfiability of the corresponding subformula $\fml{F}' \subseteq \fml{F}$. If it is
satisfiable, $\fml{F}'$ represents an MSS of $\fml{F}$, and the clause $I \setminus P$ is added to $\fml{Q}$, preventing the algorithm from generating any subset of the MSS (superset of the MCS) again. Otherwise, if $\fml{F}'$ is unsatisfiable, it is reduced to an MUS $\fml{M}$, which is blocked adding to $\fml{Q}$ a clause made of the variables in $I$ associated with $\fml{M}$ with negative polarity. This way, no superset of $\fml{M}$ will be generated. Algorithm \ref{alg:emus} is guaranteed to find all MUSes and MCSes of $\fml{F}$, in a number of iterations that corresponds to the sum of the number of MUSes and MCSes.

This paper considers the concrete problem of enumerating the
group-MUSes of an unsatisfiable Horn formula. As highlighted earlier,
and as discussed later in the paper, enumeration of the group-MUSes of
unsatisfiable Horn formulae finds important applications in axiom
pinpointing for the \elplain family of DLs, including \elplus.
It should be observed that the difference between the enumeration of
plain MUSes of Horn formulae and the enumeration of group-MUSes is
significant.
First, enumeration of group-MUSes of Horn formulae cannot be achieved
in total polynomial time, unless $\tn{P}=\tn{NP}$. This is an
immediate consequence from the fact that axiom pinpointing for the
\elplain family of DLs cannot be achieved in total polynomial time,
unless $\tn{P}=\tn{NP}$~\cite{baader-ki07}, and that axiom pinpointing
for the \elplain family of DLs can be reduced in polynomial time to
group-MUS enumeration of Horn formulae~\cite{ams-corr15}.
Second, enumeration of MUSes of Horn formulae can be achieved in total
polynomial time (actually with polynomial delay)~\cite{penaloza-kr10}.

Given the above, a possible approach for enumerating group-MUSes of
Horn formulae is to use an existing solution, either based on explicit
or implicit minimal hitting set dualization. For example, the use of
explicit minimal hitting dualization was recently proposed in ~\cite{ams-corr15}.
Alternatively, either \emus~\cite{pms-aaai13} or the different versions
of MARCO~\cite{liffiton-cpaior13,lpmms-cj15} could be used, as also
pointed out in~\cite{mp-tr15}.

This paper opts instead to exploit the implicit minimal hitting set
dualization approach~\cite{liffiton-cpaior13,pms-aaai13,lpmms-cj15},
but develops a solution that is specific to the problem formulation.
This solution is described in the next section.

\section{Algorithm for Group-MUS Enumeration in Horn Formulae}
\label{sec:hegmus}

This section describes \hgmus, a novel and efficient group-MUS
enumerator for Horn formulae based on implicit minimal hitting set
dualization.
In this section, $\fml{H}$ denotes the group of clauses $\fml{G}_0$,
i.e.\ the background clauses. Moreover, $\fml{I}$ denotes the set of
(individual) groups of clauses, with
$\fml{I}=\{\fml{G}_1,\ldots,\fml{G}_k\}$.
So, the unsatisfiable group-Horn formula corresponds to
$\fml{F}=\fml{H}\cup\fml{I}$.
Also, in this section, the formula $\fml{Q}$ shown
in~\autoref{alg:emus} is defined on a set of variables associated to
the groups in $\fml{I}$.
For the problem instances considered
later in the paper (obtained from axiom pinpointing for the \elplain
family of DLs), each group of clauses contains a single unit clause.
However, the algorithm would work for arbitrary groups of clauses.

\subsection{Organization}

The high-level organization of \hgmus mimics that of \emus/MARCO
(see~\autoref{alg:emus}), with a few essential differences.
First, the satisfiability testing step (because it operates on Horn
formulae) uses the dedicated linear time algorithm
LTUR~\cite{minoux-ipl88}. LTUR can be viewed as one-sided unit
propagation, since only variables assigned value 1 are propagated.
Moreover, the simplicity of LTUR enables very efficient
implementations, that use adjacency lists for representing clauses
instead of the now more commonly used watched literals.
Second, the problem formulation motivates using a dedicated MUS
extraction algorithm, which is shown to be more effective in this
concrete case than other well-known approaches~\cite{blms-aicomm12}. 
Third, we also highlight important aspects of the \emus/MARCO implicit
minimal hitting set dualization approach, which we claim have been
overlooked in earlier work~\cite{vescovi-phd11,sebastiani-tr15}.

\subsection{Computing Maximal Models}

\begin{algorithm}[t]
{\small 
\DontPrintSemicolon
\SetAlgoNoLine
\LinesNumbered
\SetFillComment
\SetKw{KwNot}{not\xspace}
\SetKw{KwAnd}{and\xspace}
\SetKw{KwOr}{or\xspace}
\SetKwData{false}{{\small false}}
\SetKwData{true}{{\small true}}
\SetKwData{st}{\small{\sl st}}
\SetKwFunction{SAT}{{\normalsize SAT}}
\SetKwFunction{InitAssign}{InitialAssignment}
\SetKwFunction{UpdateSAT}{UpdateSatClauses}
\SetKwFunction{PickLit}{SelectLiteral}
\SetKwFunction{DCall}{DCall}
\SetKwFunction{Compute_MxM}{{\sc els}} %
\LinesNotNumbered
\LinesNumbered
\KwIn{$\fml{Q}$ a CNF formula}
\KwOut{$(st, P)$: with $st$ a Boolean and $P$ an MxM (if it exists)} 
\smallskip
\SetAlgoVlined
  $(P, U, B) \gets (\{ \{x\} ~|~ \lnot x \notin L(\fml{Q}) \}, \{ \{x\} ~|~ \lnot x \in L(\fml{Q}) \}, \emptyset) $\;
  $(st, P, U) \gets \InitAssign(\fml{Q} \cup P )$ \;
  \lIf{\KwNot $st$} { \Return{$(\false, \emptyset)$} } 
  
   \While{$U \neq \emptyset$}{
   	$l \gets \PickLit(U)$ \;
  
   	$(st, \mu) \gets \SAT(\fml{Q} \cup P \cup B \cup \{l\})$\;
   	\leIf{$st$}
   	{
   		$(P, U) \gets \UpdateSAT(\mu, P, U) $\;
   
   	}{
   		$(U, B) \gets ( U \setminus \{l\}  ,B \cup \{\lnot l\})$
   	}
   }
  \Return{$(\true, P)$} \tcp*[r]{P is an MxM of $\fml{Q}$} 
  \SetAlgoShortEnd
}
\caption{Computation of Maximal Models \label{alg:mxm}}
\end{algorithm}

The use of maximal models for computing either MCSes of a formula or a
set of clauses that contain an MUS was proposed in earlier
work~\cite{pms-aaai13}, which exploited SAT with preferences for
computing maximal
models~\cite{giunchiglia-ecai06,giunchiglia-aicomm13}.
The use of SAT with preferences for computing maximal models is also
exploited in related work~\cite{sebastiani-cade09,sebastiani-tr15}.

Computing maximal models of a formula $\fml{Q}$ can be reduced to the problem of 
extracting an MSS of a formula $\fml{Q}'$ \cite{mshjpb-ijcai13}, where the clauses of $\fml{Q}$ are hard and, for each variable $x_i \in V(\fml{Q})$, it includes a unit soft clause $c_i \equiv \{x_i\}$.
Also, recent work ~\cite{mshjpb-ijcai13,mazure-aaai14,bacchus-aaai14b,mpms-ijcai15} has shown that state-of-the-art MCS/MSS computation approaches outperform SAT with preferences.
\textsc{HgMUS} uses a dedicated algorithm based on the LinearSearch MCS extraction algorithm \cite{mshjpb-ijcai13}, due to its good performance in MCS enumeration. Since all soft clauses are unit, it can also be related with the novel Literal-Based eXtractor algorithm \cite{mpms-ijcai15}. Shown in~\autoref{alg:mxm}, it relies on making successive calls to a SAT solver. It maintains three sets of literals: $P$, an under-approximation of an MxM (i.e. positive literals s.t. $\fml{Q} \cup P \nentails \bot$), $B$, with negative literals $\lnot l$ such that $\fml{Q} \cup P \cup \{l\} \entails \bot$ (i.e. \textit{backbone literals}), and $U$, with the remaining set of positive literals to be tested. Initially, $P$ and $U$ are initialized from a model $\mu \entails \fml{Q}$, $P$ ($U$) including the literals appearing with positive (negative) polarity in $\mu$. Then, iteratively, it tries to extend $P$ with a new literal $l\in U$, by testing the satisfiability of $\fml{Q} \cup P \cup B \cup \{l\}$. If it is satisfiable, all the literals in $U$ satisfied by the model (including $l$) are moved to $P$. Otherwise, $l$ is removed from $U$ and $\lnot l$ is added to $B$. This algorithm has a query complexity of $\fml{O}(|V(\fml{Q})|)$. 

\autoref{alg:mxm} integrates a new technique, which consists in pre-initializating $P$ with the pure positive literals appearing in $\fml{Q}$ and $U$ with the remaining ones (line 1), and then requiring the literals of $P$ to be satisfied by the initial assignment (line 2). It can be easily proved that these pure literals are included in all MxMs of $\fml{Q}$, so a number of calls to the SAT solver could be avoided. Moreover, the SAT solver will never branch on these variables, easing the decision problems.
This technique is expected to be effective in \textsc{HgMUS}. Note that, in this context, $\fml{Q}$ is made of two types of clauses: positive clauses blocking MCSes of the Horn formula, and negative clauses blocking MUSes. So, with this technique, the computation of MxMs is restricted to the variables representing groups appearing in some MUS of the Horn formula.\footnote{SATPin~\cite{mp-tr15} also exploits this insight of \textit{relevant} variables, but not in the computation of MxMs, as SATPin does not compute MxMs.}

\subsection{Adding Blocking Clauses}

One important aspect of \hgmus are the blocking clauses created and
added to the formula $\fml{Q}$ (see~\autoref{alg:emus}).
These follow what was first proposed in \emus~\cite{pms-aaai13} and
MARCO~\cite{liffiton-cpaior13,lpmms-cj15}.
For each MUS, the blocking clause consists of a set of negative
literals, requiring at least one of the clauses in the MUS {\em not}
to be included in future selected sets of clauses. For each MCS, the
blocking clause consists of a set of positive literals, requiring at
least one of the clauses in the MCS to be included in future selected
sets of clauses.
The way MCSes are handled is essential to prevent that MCS and sets
containing the same MCS to be selected again.
Although conceptually simple, it can be shown that existing approaches
may not guarantee that supersets of MCSes (or subsets of the MSSes)
are not selected. As argued later, this is the case with
\elsat~\cite{vescovi-phd11,sebastiani-tr15}.

\subsection{Deciding Satisfiability of Horn Formulae}

It is well-known that Horn formulae can be decided in linear
time~\cite{gallier-jlp84,itai-jlp87,minoux-ipl88}.
\hgmus implements the LTUR algorithm~\cite{minoux-ipl88}. 
There are important reasons for this choice. First, LTUR is expected
to be more efficient than plain unit propagation, since only variables
assigned value 1 need to be propagated. Second, most implementations
of unit propagation in CDCL SAT solvers (i.e.\ that use watched
literals) are not guaranteed to run in linear time~\cite{gent-jair13};
this is for example the case will {\em all} implementations of
Minisat~\cite{een-sat03} and its variants, for which unit propagation
runs in worst-case quadratic time. As a result, using an off-the-shelf
SAT solver and exploiting only unit propagation (as is done for
example in earlier
work~\cite{sebastiani-cade09,sebastiani-tr15,mp-tr15}) is unlikely to
be the most efficient solution.
Besides the advantages listed above, the use of a linear time
algorithm for deciding the satisfiability of Horn formulae turns out
to be instrumental for MUS extraction, as shown in the next section.
In order to use LTUR for MUS extraction, an incremental version has
been implemented, which allows for the incremental addition of clauses
to the formula and incremental identification of variables assigned
value 1. Clearly, the amortized run time of LTUR, after adding $m =
|\fml{F}|$ clauses, is $\fml{O}(||\fml{F}||)$, with $||\fml{F}||$ the
number of literals appearing in $\fml{F}$.

\subsection{MUS Extraction in Horn Formulae} \label{ssec:himus}

For arbitrary CNF formulae, a number of approaches exist for MUS
extraction, with the most commonly used one being the deletion-based
approach~\cite{bakker-ijcai93,blms-aicomm12}, but other alternatives
include the QuickXplain algorithm~\cite{junker-aaai04} and the more
recent Progression algorithm~\cite{msjb-cav13}. It is also well-known
and generally accepted that, due to its query complexity, the 
insertion-based algorithm~\cite{puget-ecai88} for MUS extraction is in
practice not competitive with existing
alternatives~\cite{blms-aicomm12}.

\begin{algorithm}[t]
{\small 
\DontPrintSemicolon
\SetAlgoNoLine
\LinesNumbered
\SetFillComment
\SetKw{KwNot}{not\xspace}
\SetKw{KwAnd}{and\xspace}
\SetKw{KwOr}{or\xspace}
\SetKwData{false}{{\small false}}
\SetKwData{true}{{\small true}}
\SetKwData{st}{\small{\sl st}}
\SetKwFunction{prop}{LTUR\_prop}
\SetKwFunction{undo}{LTUR\_undo}
\SetKwFunction{selcl}{SelectRemoveClause}
\SetKw{KwBreak}{break\xspace}

\KwIn{$\fml{H}$, denotes the $\fml{G}_0$ clauses; $\fml{I}$, denotes
  the set of (individual) group clauses}
\KwOut{$\fml{M}$, denotes the computed MUS}
\smallskip
\SetAlgoVlined
$(\fml{M},c_r)\gets(\fml{H},0)$ \;
$\prop(\fml{M},\fml{M})$ \tcp*[r]{Start by propagating $\fml{G}_0$ clauses}
\While{\true}{
  \If{$c_r>0$}{
    $\fml{M}\gets\fml{M}\cup\{c_r\}$ \tcp*[r]{Add transition clause $c_r$ to $\fml{M}$}
    \If{\KwNot $\prop(\fml{M}, \{c_r\})$}{
      $\undo(\fml{M},\fml{M})$ \;
      \Return $\fml{M}\setminus\fml{H}$ \tcp*[r]{Remove $\fml{G}_0$
        clauses from computed MUS}
    }
  }
  $\fml{S}\gets\emptyset$ \;
  \While{\true}{
    $c_r \gets \selcl(\fml{I})$ \tcp*[r]{Target transition clause}
    $\fml{S}\gets\fml{S}\cup\{c_r\}$\;
    \If{\KwNot $\prop(\fml{M}\cup\fml{S},\{c_r\})$}{
      $\fml{I}\gets\fml{S}\setminus\{c_r\}$ 
      \tcp*[r]{Update working set of groups} 
      $\undo(\fml{M},\fml{S})$ \;
      \KwBreak \tcp*[r]{$c_r$ represents a transition clause}
    }
  }
}
}
  \caption{Insertion-based~\cite{puget-ecai88} MUS extraction using
    LTUR~\cite{minoux-ipl88}} \label{alg:himus}
\end{algorithm}

Somewhat surprisingly, this is not the case with Horn formulae when
(an incremental implementation of) the LTUR algorithm is used.
A modified insertion-based MUS extraction algorithm that exploits LTUR
is shown in~\autoref{alg:himus}.
\prop propagates the consequences of adding some new set of clauses,
given some existing incremental context.
\undo unpropagates the consequences of adding some set of clauses (in
order), given some existing incremental context.
The organization of the algorithm mimics the standard insertion-based
MUS extraction algorithm~\cite{puget-ecai88}, but the use of the
incremental LTUR yields run time complexity that improves over other
approaches.
Consider the operation of the standard insertion-based
algorithm~\cite{puget-ecai88}, in which clauses are iteratively added
to the working formula. When the formula becomes unsatisfiable, a
{\em transition clause}~\cite{blms-aicomm12} has been identified,
which is then added to the MUS being constructed.
The well-known query complexity of the insertion-based algorithm is
$\fml{O}(m\times k)$ where $m$ is the number of clauses and $k$ is the
size of a largest MUS.
Now consider that the incremental LTUR algorithm is used. To find the
first transition clause, the amortized run time is
$\fml{O}(||\fml{F}||)$. Clearly, this holds true for {\em any}
transition clause, and so the run time of MUS extraction with the LTUR
algorithm becomes $\fml{O}(|\fml{M}|\times||\fml{F}||)$, where
$\fml{M}\subseteq\fml{I}$ is a largest MUS.
\autoref{alg:himus} highlights the main differences with respect to a
standard insertion-based MUS extraction algorithm.
In contrast, observe that for a deletion-based algorithm the run time
complexity will be $\fml{O}(|\fml{I}|\times||\fml{F}||)$. In
situations where the sizes of MUSes are much smaller than the number
of groups in $\fml{I}$, this difference can be significant. As a
result, when extracting MUSes from Horn formulae, and when using a
polynomial time incremental decision procedure, an insertion-based
algorithm should be used instead of other more commonly used
alternatives.

\section{Comparison with Existing Alternatives} \label{sec:relw}

This section compares \hgmus with the group-MUS enumerators used in
\elsat~\cite{sebastiani-cade09,sebastiani-tr15},
and SATPin~\cite{mp-tr15}.
The experimental comparison with these enumerators as well as other
approaches for axiom pinpointing for the \elplain family of DLs is
provided in~\autoref{sec:res}.

\subsection{\elsat}

The best known SAT-based approach for axiom pinpointing is
\elsat~\cite{sebastiani-cade09,vescovi-phd11,sebastiani-tr15}.
\elsat is composed of two main phases. The first phase compiles the
axiom pinpointing problem to a Horn formula. 
The second phase enumerates the so-called MinAs, and corresponds to
group-MUS enumeration for this Horn formula~\cite{ams-corr15}..
Although existing references emphasize the enumeration of MinAs (MUSes) using
an AllSAT approach (itself inspired by an AllSMT
approach~\cite{nieuwenhuis-cav06}), the connection with MUS
enumeration is immediate~\cite{ams-corr15}. More importantly, \elsat
shares a number of similarities with implicit minimal hitting set
dualization, but also crucial differences, which we now analyze.

Similar to \emus, \elsat selects subformulae of an unsatisfiable Horn
formula. This is achieved with a SAT solver that always assigns
variables value 1 when branching~\cite{sebastiani-tr15}. This
corresponds to solving SAT with
preferences~\cite{giunchiglia-ecai06,giunchiglia-aicomm13}, and so it
corresponds to computing a maximal model, inasmuch the same way as
\emus operates.

In \elsat, the approach for deciding the satisfiability of Horn
subformulae is based on running the unit propagation engine of a CDCL 
SAT solver. As explained earlier, this can be inefficient when
compared with the dedicated LTUR algorithm for Horn
formulae~\cite{minoux-ipl88}.
Moreover, in \elsat, MUSes are extracted with what can be viewed as a
deletion-based algorithm~\cite{bakker-ijcai93,blms-aicomm12}. Although
more efficient alternatives are suggested, none is as asymptotically
as efficient as the dedicated algorithm proposed
in~\autoref{ssec:himus}.

Finally, the most important drawback is the blocking of sets of
clauses that do not contain an MUS/MinA. In our setting of implicit
minimal hitting set dualization, this represents one MCS. The approach
used in \elsat consists of creating a blocking clause solely based on
the decision variables (which are {\em always} assigned value
1)~\cite{vescovi-phd11,sebastiani-tr15}\footnote{The clause learning
  mechanism used in \elsat is detailed in~\cite[page 17,
    first paragraph]{sebastiani-tr15}.}. Thus, the learned clauses,
although blocking one MCS (and corresponding MSS), do {\em not} block 
supersets of MCSes (and the corresponding subsets of the MSSes). This
can result in exponentially more iterations than necessary, and
explains in part the poor performance of \elsat in practice.
It should be further observed that this drawback becomes easier to
spot once the problem is described as MUS enumeration by implicit
minimal hitting set dualization.

\subsection{SATPin}

SATPin~\cite{mp-tr15} represents a recent SAT-based alternative for
axiom pinpointing for the \elplain family of DLs, that focuses on
optimizing the low-level implementation details of the CDCL SAT
solver, including the use of incremental SAT solving. As indicated
above, \hgmus opts to revisit instead the LTUR~\cite{minoux-ipl88}
algorithm from the late 80s, since it is guaranteed to run in linear
time for Horn formulae, and can be implemented with small overhead.
The SATPin approach is presented in terms of iteratively computing
prime implicants. However, computing a prime implicant is tightly
related with extracting an MUS~\cite{bradley-fmcad07}. As a result,
some aspects of the organization of SATPin can be related with those
of \elsat, namely the procedure for extracting MUSes/MinAs.
Although the actual enumeration of candidate sets is not detailed
in~\cite{mp-tr15}, the description of SATPin suggests the use of model
enumeration with some essential pruning techniques.

\section{Experimental Results} \label{sec:res}

This section evaluates group MUS enumerators for Horn formulae
obtained from axiom pinpointing problems for the \elplain family of
DLs, particularly applied to medical ontologies. 
A set of standard benchmarks is considered. These have been used in
earlier
work,
e.g.~\cite{baader-ijcar06,sebastiani-cade09,ludwig-ore14,ams-corr15,mp-tr15}.

Since all experiments consist of converting axiom pinpointing problems
into group MUS enumeration problems, the tool that uses \hgmus as its
back-end is named EL2MUS. Thus, in this section, the results for
EL2MUS illustrate the performance of the group-MUS enumerator
described in this paper.

\subsection{Experimental Setup}

Each considered instance represents the problem of explaining a particular 
subsumption relation (query) entailed in a medical ontology. Four
medical ontologies\footnote{GENE, GALEN and NCI ontologies are freely
  available at
  \url{http://lat.inf.tu-dresden.de/~meng/toyont.html}. The SNOMED~CT 
  ontology was requested from IHTSDO under a nondisclosure license
  agreement.} are considered: GALEN~\cite{rector-97}, GENE~\cite{ashburner-ng00},
NCI~\cite{sioutos-jbi07} and
SNOMED~CT~\cite{spackman-amia97}.  For GALEN, we consider two
variants: FULL-GALEN and NOT-GALEN. The most important ontology is
SNOMED~CT and, due to its huge size, it also produces the hardest
axiom pinpointing instances.
For each ontology (including the GALEN variants) 100 queries are considered; 50 random (which are expected to be easier) and 50 sorted (expected to have a large number of minimal explanations) queries. 
So, there are 500 queries in total.
Given an ontology and a subsumption query, the encoding proposed in~\cite{sebastiani-cade09,sebastiani-tr15} produces a Horn formula and a set of axioms (variables) which may be responsible for the subsumption relation. This can be transformed into a group-MUS enumeration problem where the original Horn formula forms group-0 and each axiom constitutes a group containing only a unit clause.

Two different experiments were considered by applying two 
different simplification techniques to the problem instances,
both of which were proposed in~\cite{sebastiani-tr15}. The first one uses 
the Cone-Of-Influence (COI) reduction. 
These are reduced instances in both the size of the 
Horn formula and the number of axioms, but are still quite large. 
Similar techniques are exploited in
related work~\cite{baader-ijcar06,ludwig-ore14,mp-tr15}. 
The second one considers the more effective reduction technique 
(which we refer to as x2), consisting in applying the COI technique,
re-encoding the Horn formula into a reduced ontology, and encoding
this ontology again into a Horn formula. This results in small Horn
formulae, which will be useful to evaluate the algorithms when there
are a large number of MUSes/MCSes.

The experiments compare EL2MUS to different algorithms, namely
\elsat~\cite{sebastiani-cade09,sebastiani-tr15},
CEL~\cite{baader-ijcar06}, \just~\cite{ludwig-ore14} 
and SATPin~\cite{mp-tr15}.
\elsat~\cite{sebastiani-tr15} has been shown to outperform
CEL~\cite{baader-ijcar06}, whereas SATPin~\cite{mp-tr15} has been
shown to outperform the MUS enumerator MARCO~\cite{lpmms-cj15}.

The comparison with CEL and \just imposes a number of constraints.
First, CEL only computes 10 MinAs, so all comparisons with
CEL only consider reporting the first 10 MinAs/MUSes. Also, CEL uses
a simplification technique similar to COI, so CEL is considered in the first
 experiments.
Second, \just operates on selected subsets of \elplus, i.e.\ the
description logic used in most medical ontologies. As a result, all
comparisons with \just consider solely the problem instances for which
\just can compute correct results. \just accepts the simplified x2 ontologies, 
so it is considered in the second experiments.
The comparison with these tools is presented at the end of the section.

EL2MUS interfaces the SAT solver Minisat 2.2~\cite{een-sat03} for computing maximal models. All the experiments were performed on a Linux cluster (2 GHz) and the 
algorithms were given a time limit of 3600s and a memory limit of 4
GB.

\begin{figure}[t]
  \centering
  \includegraphics[scale=0.4575]{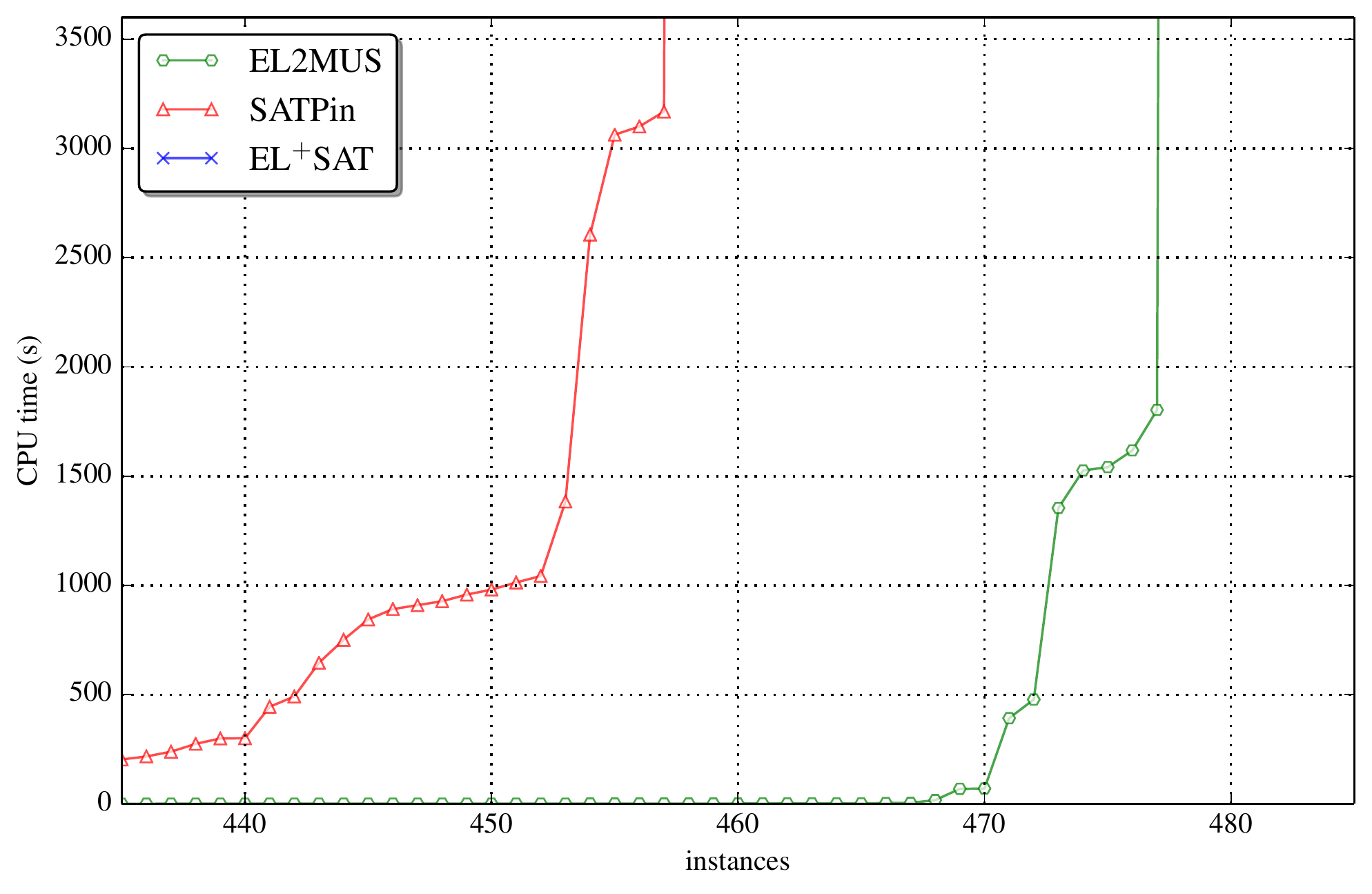} 
  \caption{Cactus plot comparing \elsat, SATPin and
    EL2MUS on the COI instances} \label{fig:cactus-coi}
\end{figure}

\subsection{COI Instances}

\autoref{fig:cactus-coi} summarizes the results for \elsat, 
SATPin and EL2MUS.
\begin{figure}[t]
  \captionsetup[subfigure]{aboveskip=-1pt,belowskip=0pt}
  \captionsetup[figure]{belowskip=-20pt}
  \centering
  \begin{subfigure}[c]{0.475\textwidth}
    \includegraphics[scale=0.45]{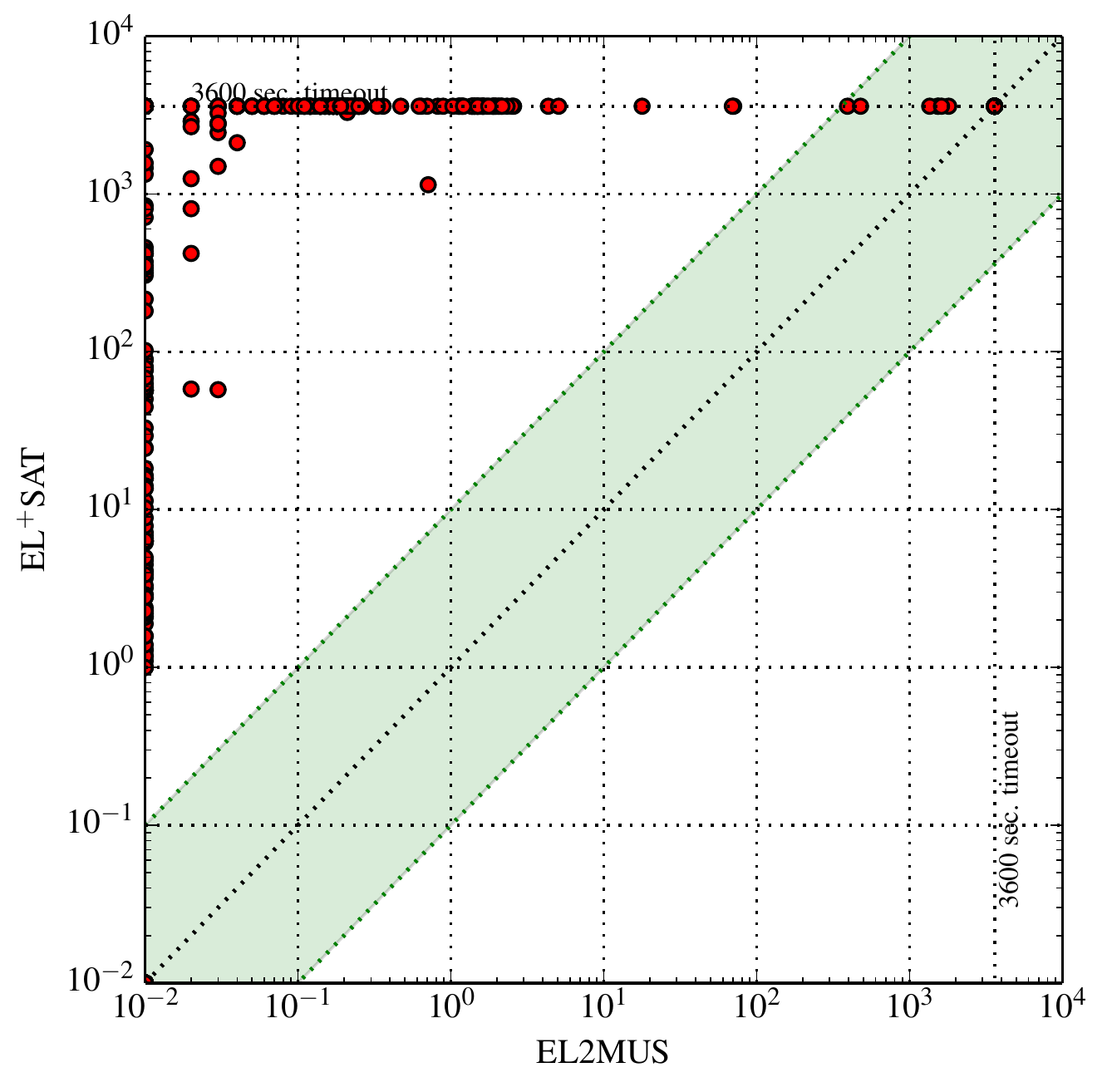}
    \caption{Comparison with \elsat}
  \end{subfigure}
  \quad
  \begin{subfigure}[c]{0.475\textwidth}
    \includegraphics[scale=0.45]{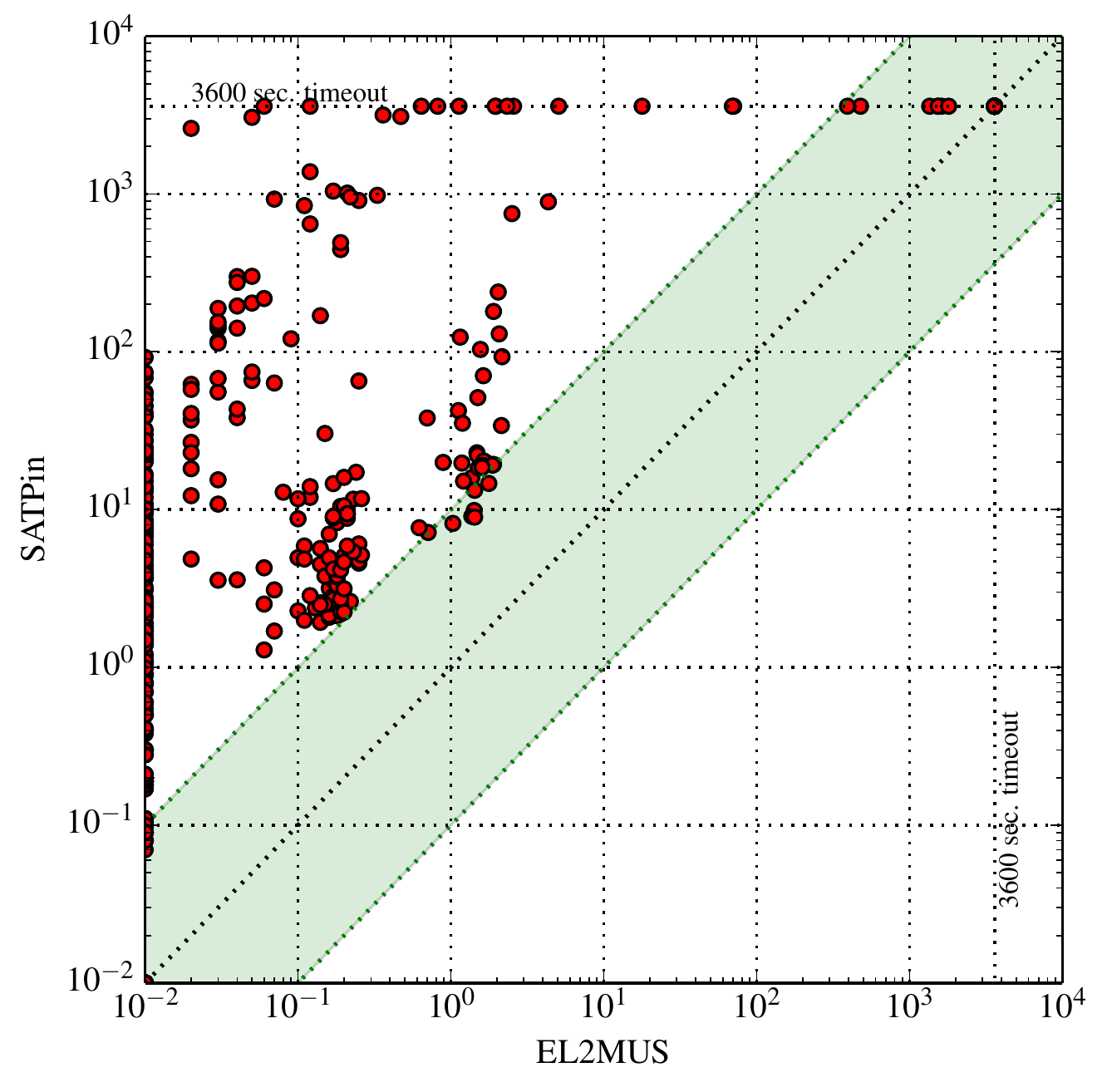}
    \caption{Comparison with SATPin}
  \end{subfigure}

  \centering
  \begin{subfigure}[c]{0.475\textwidth}
    \includegraphics[scale=0.45]{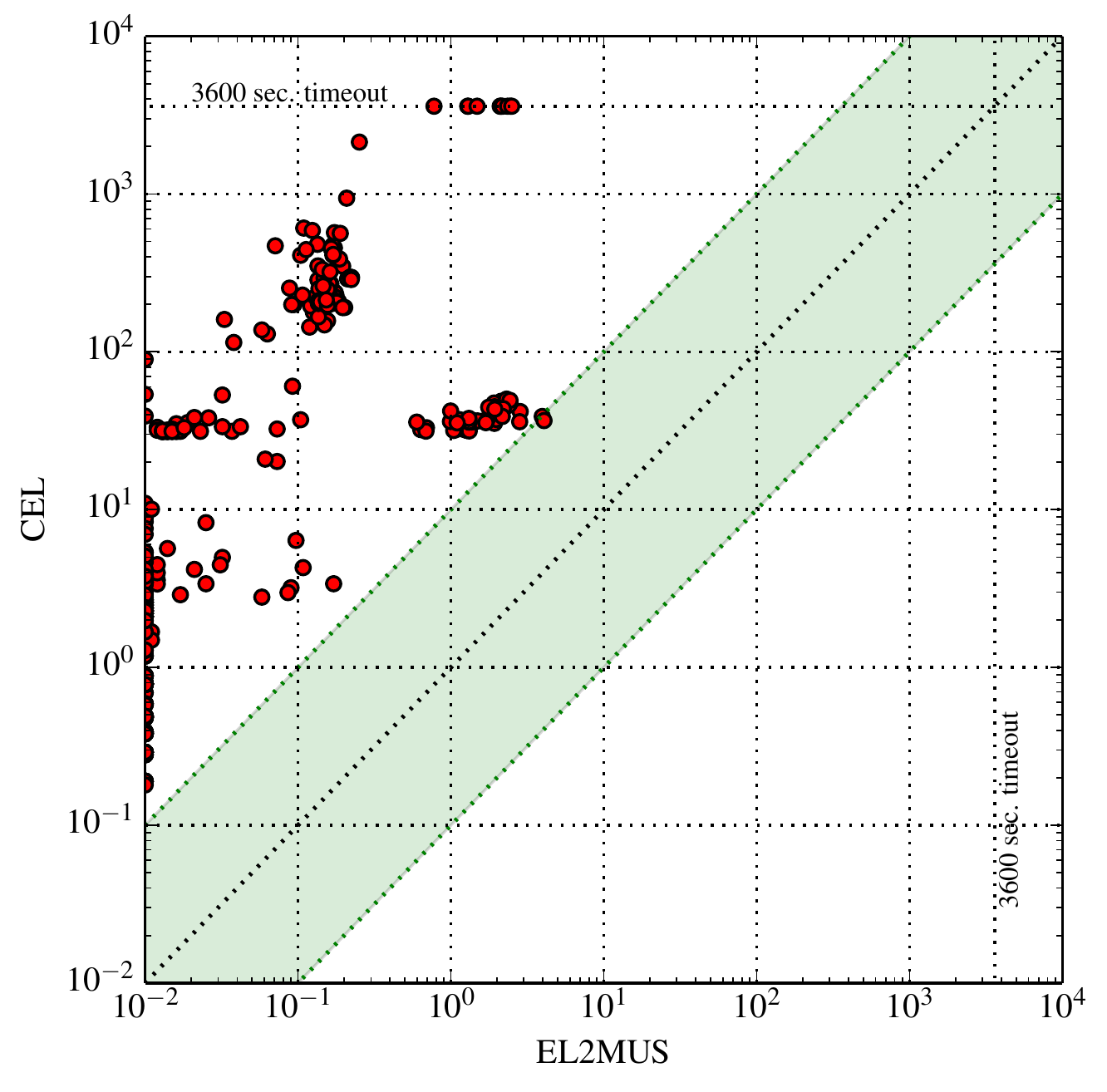}
    \caption{Comparison with CEL} \label{fig:celcoi}
  \end{subfigure}
  \quad\quad
  \begin{subfigure}[c]{0.45\textwidth}
    \centering
    \scalebox{0.9}{
    	\renewcommand{\arraystretch}{1.25}
	\begin{tabular}{|c||c|c|} \hline
	\% wins& \elsat  & SATPin  \\ \hline\hline
	\elsat &  --     & 20.29\% \\ \hline\hline
	SATPin & 79.71\% &  --     \\ \hline\hline
	EL2MUS & 100.0\% & 100.0\% \\ \hline\hline
	\hline
	$> 10^1$x & 98.09\% & 96.78\% \\ \hline\hline
	$> 10^2$x & 97.55\% & 72.07\% \\ \hline\hline
	$> 10^3$x & 96.46\% & 47.75\% \\ \hline\hline
	$> 10^4$x & 74.05\% & 06.49\% \\ \hline\hline
	$> 10^5$x & 31.10\% & 00.45\% \\ \hline\hline
	\end{tabular}
    }
    \vspace*{0.625cm}
    \caption{Summary table}\label{tab:coisum}
  \end{subfigure}
  \caption{Scatter plots for COI instances} \label{fig:scatter-coi}
\end{figure}
\elsat does not show in the plot due to its poor performance. As can
be observed, 
EL2MUS terminates for more instances than
any of the other tools.
\autoref{fig:scatter-coi} shows scatter plots comparing the different
tools. As can be concluded, and with a few outliers, the performance
of EL2MUS exceeds the performance of any of the other tools by at
least one order of magnitude (and often by more). 
\autoref{tab:coisum} summarizes the results in the scatter plots,
where the percentages shown are computed for problem instances for
which at least one of the tools takes more than 0.001s. 
CEL is not shown in the table due to the special constraints mentioned above.
As can be observed, EL2MUS outperforms any of the other tools in all
of the problem instances and, for many cases, with two or more orders
of magnitude improvement.

\begin{figure}[t]
  \centering
  \begin{subfigure}[c]{0.495\textwidth}
    \includegraphics[scale=0.45]{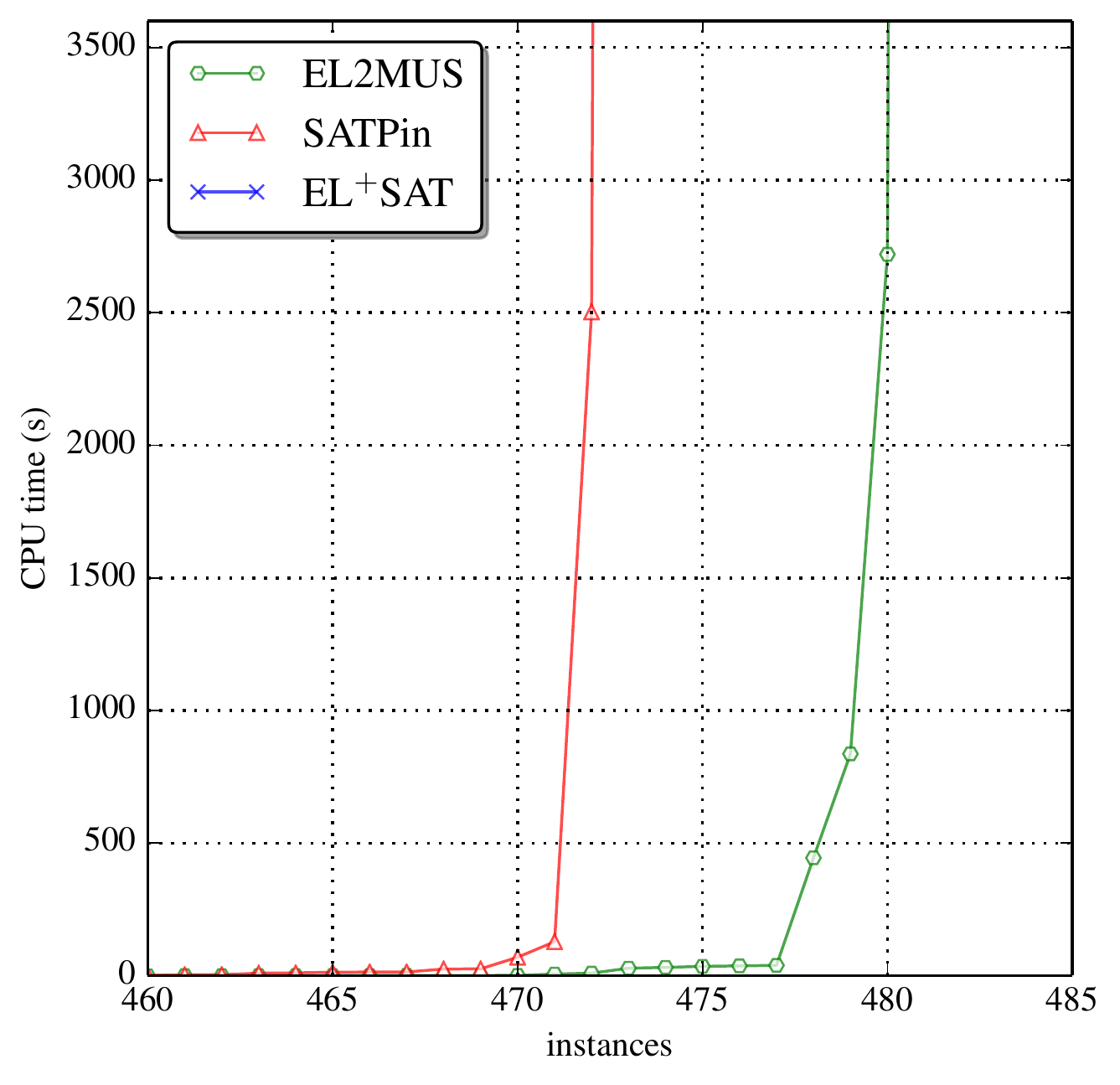}
    \caption{Solved instances}\label{fig:cactus-x2}
  \end{subfigure}
  \begin{subfigure}[c]{0.495\textwidth}
    \includegraphics[scale=0.45]{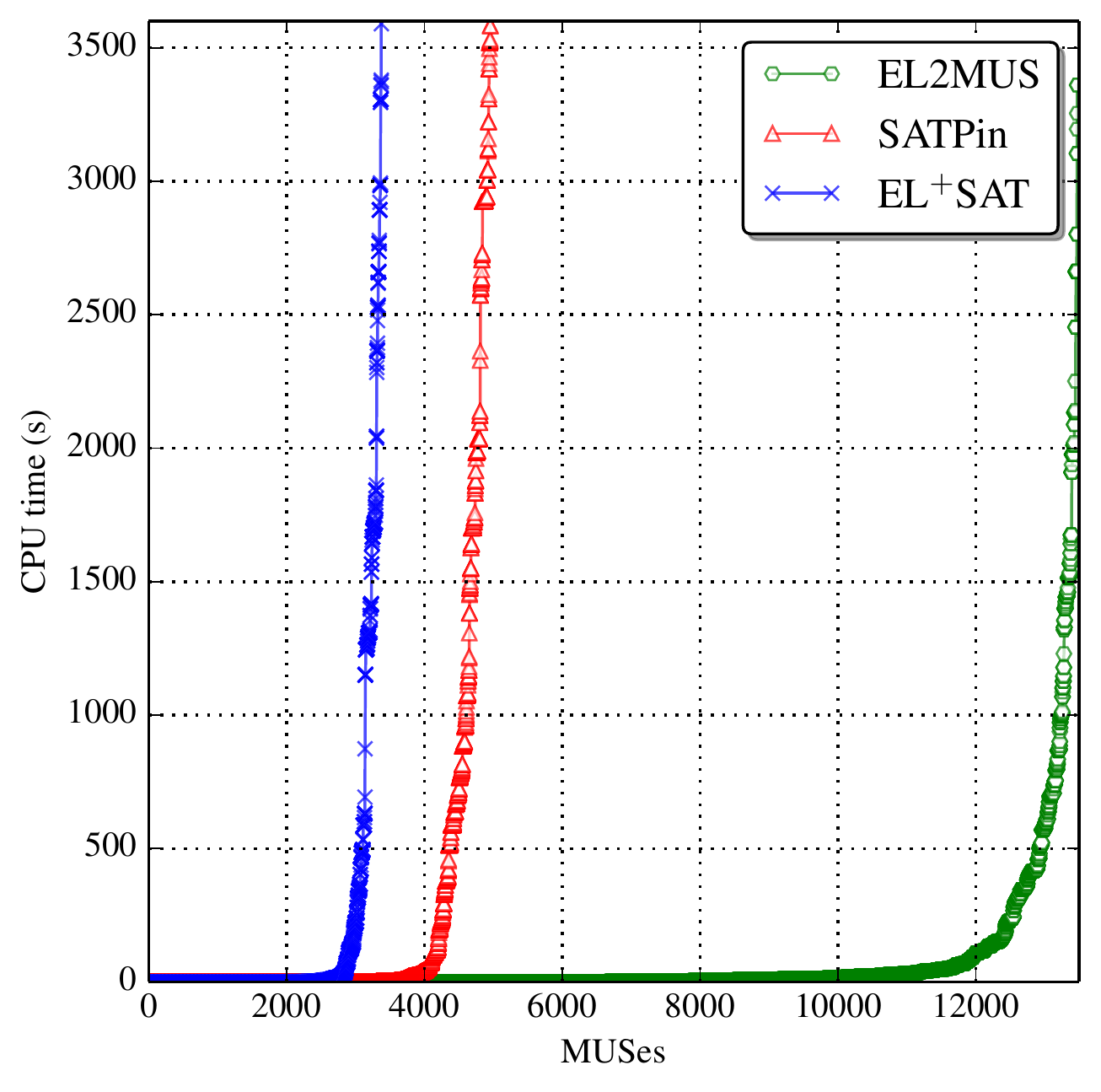}
    \caption{Reported MUSes}\label{fig:muses-x2}
  \end{subfigure}
  \caption{Cactus plots comparing \elsat, SATPin and EL2MUS on the x2 instances} \label{fig:x2cactus}
\end{figure}

\begin{figure}[t]
  \captionsetup[subfigure]{aboveskip=-1pt,belowskip=0pt}
  \captionsetup[figure]{belowskip=-20pt}
  \centering
  \begin{subfigure}[c]{0.475\textwidth}
    \includegraphics[scale=0.45]{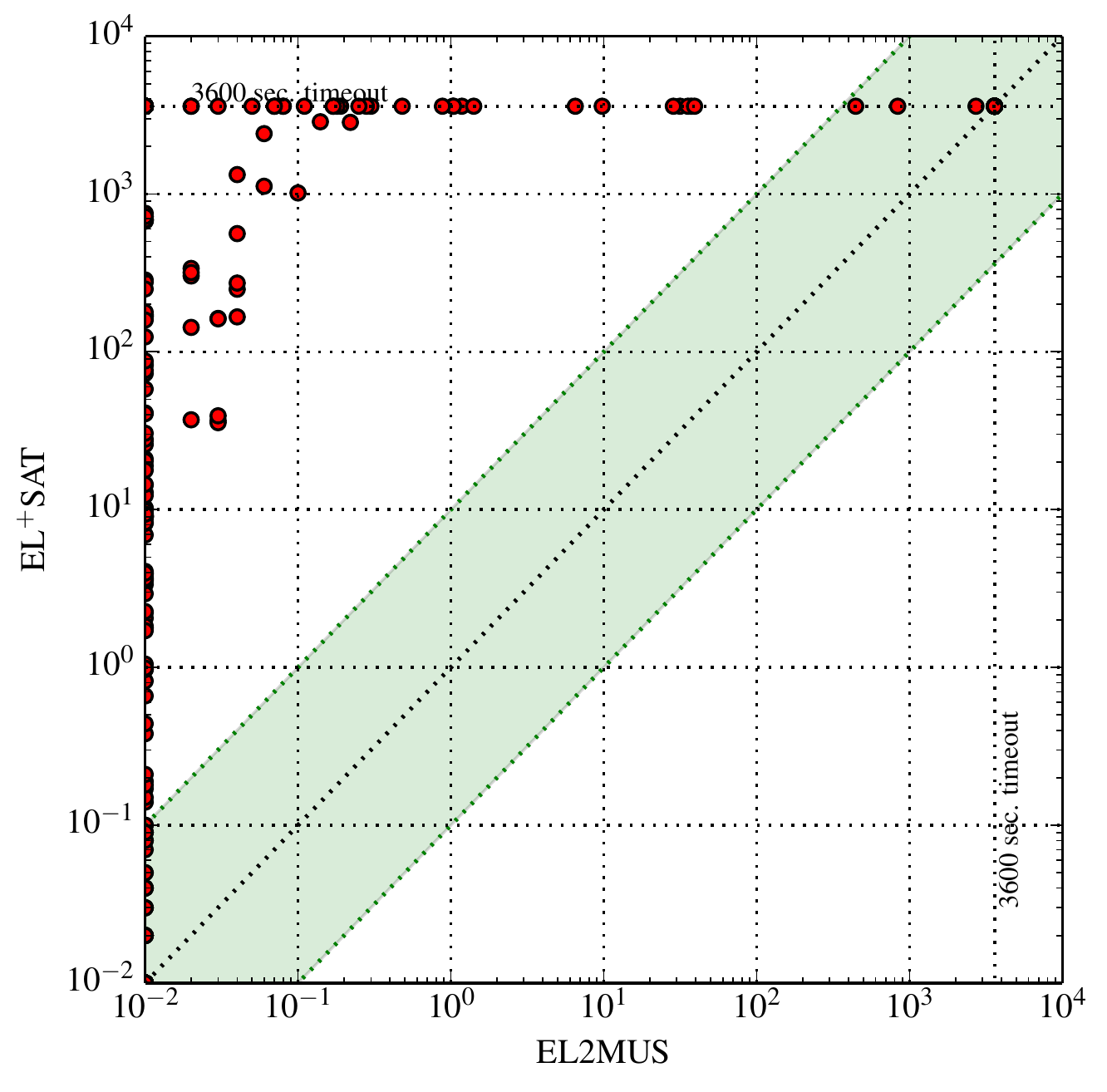}
    \caption{Comparison with \elsat}
  \end{subfigure}
  \quad
  \begin{subfigure}[c]{0.475\textwidth}
    \includegraphics[scale=0.45]{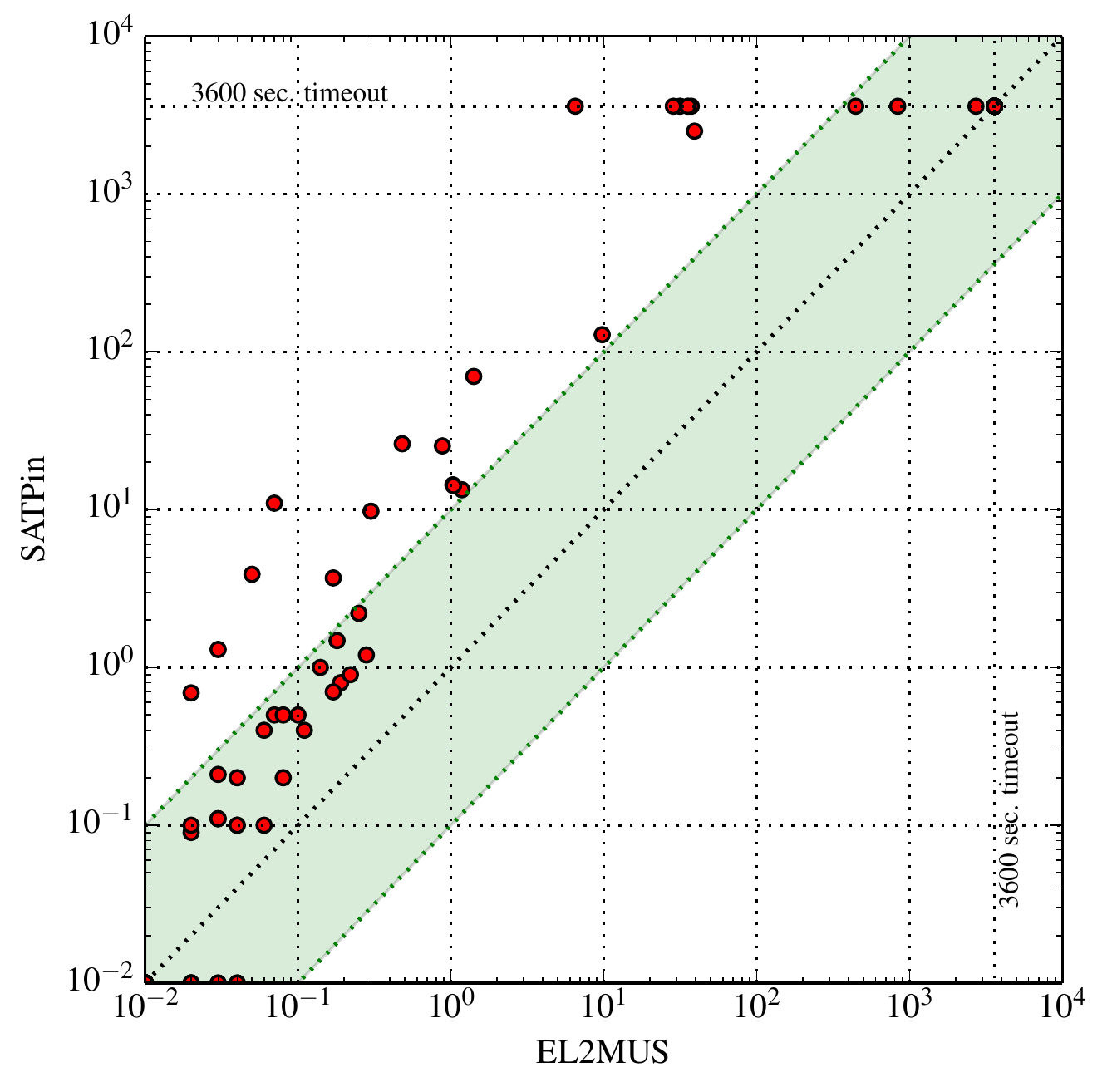}
    \caption{Comparison with SATPin}
  \end{subfigure}

  \centering
  \begin{subfigure}[c]{0.475\textwidth}
    \includegraphics[scale=0.45]{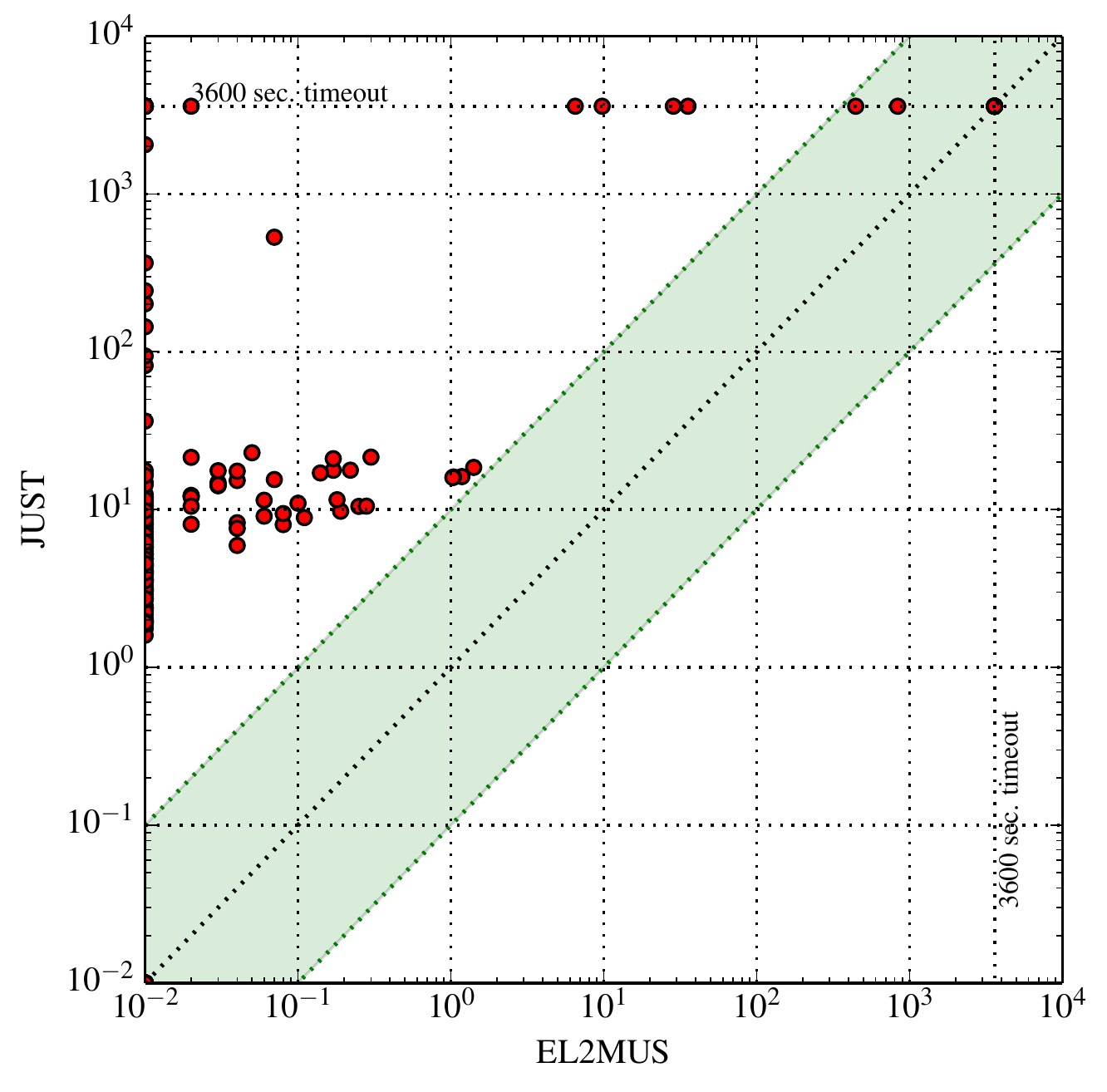}
    \caption{Comparison with \just} \label{fig:justx2}
  \end{subfigure}
  \quad\quad
  \begin{subfigure}[c]{0.45\textwidth}
    \centering
    \vspace*{0.60 cm}
    \scalebox{0.9}{
    	\renewcommand{\arraystretch}{1.25}
	\begin{tabular}{|c||c|c|} \hline
	\% wins& \elsat  & SATPin   \\ \hline\hline
	\elsat &  --     & 00.00\% \\ \hline\hline
	SATPin & 100.0\% &  --      \\ \hline\hline
	EL2MUS & 100.0\% & 67.69\% \\ \hline\hline
	\hline
               & \elsat & SATPin  \\ \hline\hline
	\# MUSes       &  788  & 1484   \\ \hline\hline 
	$\Delta$ MUSes & 9160  & 8864   \\ \hline\hline
	\end{tabular}
    }
    \vspace*{0.675cm} 
    \caption{Summary table} \label{tab:x2sum}
  \end{subfigure}
  \caption{Scatter plots for x2 instances} \label{fig:scatter-x2}
\end{figure}

\subsection{x2 Instances}

The x2 instances are significantly simpler than the COI
instances. Thus, whereas the COI instances can serve to assess the
scalability of each approach, the x2 instances highlight the expected
performance in representative settings.
\autoref{fig:cactus-x2} summarizes the performance of the tools
\elsat, SATPin 
and EL2MUS. As with the COI instances, \elsat
does not show in the plot due to its poor performance.
Moreover, and as before in terms of terminated instances, EL2MUS
exhibits an observable performance edge.

A pairwise comparison between the different tools is summarized
in~\autoref{fig:scatter-x2}. Although not as impressive as for the COI
instances, EL2MUS still consistently outperforms all other tools.
\autoref{tab:x2sum} summarizes the results, where as before the
percentages shown are computed for problem instances for which at
least one of the tools takes more than 0.001s. Observe that, for these
easier instances, SATPin becomes competitive with EL2MUS.
Nevertheless, for instances taking more than 0.1s, EL2MUS outperforms
SATPin on 100\% of the instances. Thus, the 67.69\% shown in the table
result from instances for which both SATPin and EL2MUS take at most
0.04s.
The summary table also lists the number of computed MUSes for the 19
instances for which EL2MUS does {\em not} terminate (all of the other
tools also do not terminate for these 19 instances). EL2MUS computes
9948 MUSes in total. As can be observed from the table, the other
tools lag behind, and compute significantly fewer MUSes. 
The comparison with \elsat and
SATPin, reveals that EL2MUS computes respectively in excess of a
factor of 10 and of 5 more MUSes.

EL2MUS not only terminates on more instances than any other approach
and computes more MUSes for the unsolved instances; it also reports
the sequences of MUSes much faster. \autoref{fig:muses-x2} shows, for
each computed MUS over the whole set of instances, the time each MUS
was reported. This figure compares \elsat, SATPin and EL2MUS, as these
are the only methods able to report MUSes from the beginning. The
results confirm that EL2MUS is able to find many more MUSes in less
time than the alternatives.

These experimental results suggest that, not only is EL2MUS the best
performing axiom pinpointing tool, on both the COI and x2 problem
instances, but it is also the one that is expected to scale better for
more challenging problem instances, given the results on the COI
instances.

\subsection{Assessment of Non SAT-Based Axiom Pinpointing Tools}

\autoref{fig:celcoi} and \autoref{fig:justx2} show scatter plots comparing EL2MUS with
CEL~\cite{baader-ijcar06} and \just~\cite{ludwig-ore14}, respectively
for the COI and x2 instances\footnote{The other
  scatter plots are not shown, but the conclusions are the
  same.}.
As indicated earlier, CEL only computes 10 MinAs, and so the run times
shown are for computing the first 10 MinAs. As can be observed,
the performance edge of EL2MUS is clear, with the performance gap
exceeding 1 order of magnitude almost without exception.
Moreover, \just~\cite{ludwig-ore14} is a recent state of the art axiom
pinpointing tool for the less expressive $\fml{ELH}$ DL. Thus, not
all subsumption relations can be represented and analyzed. The results
shown are for the subsumption relations for which \just gives the
correct results. In total, 382 instances could be considered and are
shown in the plot. As before, the performance edge of EL2MUS is
clear, with the performance gap exceeding 1 order of magnitude without
exception. In this case, since  the x2 instances are in general much
simpler, the performance gap is even more significant.

\section{Conclusions \& Future Work} \label{sec:conc} 

Enumeration of group MUS for Horn formulae finds important
applications, including axiom pinpointing for the \elplain family of
DLs. Since the \elplain family of DLs is widely used for representing
medical ontologies, namely with \elplus, enumeration of group MUSes
for Horn formulae represents a promising and strategic application of
SAT technology. This includes, among others, 
SAT solvers, MCS extractors and enumerators, and MUS extractors and 
enumerators.
This paper develops a highly optimized group MUS enumerator for Horn
formulae, which is shown to extensively outperform any other existing
approach. Performance gains are almost without exception at least one
order of magnitude, and most often significantly more than that.
More importantly, the experimental results demonstrate that SAT-based
approaches are by far the most effective approaches for axiom
pinpointing for the \elplain family of DLs. When compared with other
non SAT-based approaches, the performance gains are also conclusive.

Future work will exploit integration of additional recent work on
SAT-based problem solving, e.g.\ in MCS enumeration and MUS
enumeration, to further improve performance of axiom pinpointing.

\subsubsection*{\ackname}
We thank the authors of \elsat, R.\ Sebastiani and M.\ Vescovi, for
authorizing the use of the most recent, yet unpublished, version of
their work~\cite{sebastiani-tr15}.
We thank the authors of SATPin~\cite{mp-tr15}, N.\ Manthey and
R.\ Pe\~{n}aloza, for bringing SATPin to our attention, and for
allowing us to use their tool.
We thank A.\ Biere for pointing out reference~\cite{gent-jair13}, on
the complexity of implementing unit propagation when using watched
literals.
This work is partially supported by SFI PI grant BEACON (09/IN.1/\-I2618),
by FCT grant POLARIS (PTDC/EIA-CCO/\-123051/\-2010), and by national funds
through FCT with reference UID/CEC/50021/2013.



\clearpage

\begin{thebibliography}{10}

\bibitem{ams-corr15}
M.~F. Arif and J.~Marques{-}Silva.
\newblock Towards efficient axiom pinpointing of {EL+} ontologies.
\newblock {\em CoRR}, abs/1503.08454, March 2015.
\newblock Available from \url{http://arxiv.org/abs/1503.08454}.

\bibitem{ashburner-ng00}
M.~Ashburner, C.~A. Ball, J.~A. Blake, D.~Botstein, H.~Butler, J.~M. Cherry,
  A.~P. Davis, K.~Dolinski, S.~S. Dwight, J.~T. Eppig, and et~al.
\newblock Gene ontology: tool for the unification of biology.
\newblock {\em Nature genetics}, 25(1):25--29, 2000.

\bibitem{baader-jar95}
F.~Baader and B.~Hollunder.
\newblock Embedding defaults into terminological knowledge representation
  formalisms.
\newblock {\em J. Autom. Reasoning}, 14(1):149--180, 1995.

\bibitem{baader-hdbk08}
F.~Baader, I.~Horrocks, and U.~Sattler.
\newblock Description logics.
\newblock In V.~L. Frank~van Harmelen and B.~Porter, editors, {\em Handbook of
  Knowledge Representation}, Foundations of Artificial Intelligence, chapter~3,
  pages 135 -- 179. Elsevier, 2008.

\bibitem{baader-ijcar06}
F.~Baader, C.~Lutz, and B.~Suntisrivaraporn.
\newblock {CEL} - {A} polynomial-time reasoner for life science ontologies.
\newblock In {\em IJCAR}, pages 287--291, 2006.

\bibitem{baader-jlc10}
F.~Baader and R.~Pe{\~{n}}aloza.
\newblock Axiom pinpointing in general tableaux.
\newblock {\em J. Log. Comput.}, 20(1):5--34, 2010.

\bibitem{baader-ki07}
F.~Baader, R.~Pe{\~{n}}aloza, and B.~Suntisrivaraporn.
\newblock Pinpointing in the description logic $\mathcal{EL}^{+}$.
\newblock In {\em KI}, pages 52--67, 2007.

\bibitem{baader-krmed08}
F.~Baader and B.~Suntisrivaraporn.
\newblock Debugging {SNOMED} {CT} using axiom pinpointing in the description
  logic $\mathcal{EL}^{+}$.
\newblock In {\em KR-MED}, 2008.

\bibitem{bacchus-aaai14b}
F.~Bacchus, J.~Davies, M.~Tsimpoukelli, and G.~Katsirelos.
\newblock Relaxation search: {A} simple way of managing optional clauses.
\newblock In {\em AAAI}, pages 835--841, 2014.

\bibitem{stuckey-padl05}
J.~Bailey and P.~J. Stuckey.
\newblock Discovery of minimal unsatisfiable subsets of constraints using
  hitting set dualization.
\newblock In {\em PADL}, pages 174--186, 2005.

\bibitem{bakker-ijcai93}
R.~R. Bakker, F.~Dikker, F.~Tempelman, and P.~M. Wognum.
\newblock Diagnosing and solving over-determined constraint satisfaction
  problems.
\newblock In {\em IJCAI}, pages 276--281, 1993.

\bibitem{blms-aicomm12}
A.~Belov, I.~Lynce, and J.~Marques{-}Silva.
\newblock Towards efficient {MUS} extraction.
\newblock {\em {AI} Commun.}, 25(2):97--116, 2012.

\bibitem{sat-handbook09}
A.~Biere, M.~Heule, H.~van Maaren, and T.~Walsh, editors.
\newblock {\em Handbook of Satisfiability}, volume 185 of {\em Frontiers in
  Artificial Intelligence and Applications}. {IOS} Press, 2009.

\bibitem{lozinskii-jetai03}
E.~Birnbaum and E.~L. Lozinskii.
\newblock Consistent subsets of inconsistent systems: structure and behaviour.
\newblock {\em J. Exp. Theor. Artif. Intell.}, 15(1):25--46, 2003.

\bibitem{bradley-fmcad07}
A.~R. Bradley and Z.~Manna.
\newblock Checking safety by inductive generalization of counterexamples to
  induction.
\newblock In {\em FMCAD}, pages 173--180, 2007.

\bibitem{cook-stoc71}
S.~A. Cook.
\newblock The complexity of theorem-proving procedures.
\newblock In {\em STOC}, pages 151--158, 1971.

\bibitem{puget-ecai88}
J.~L. de~Siqueira~N. and J.-F. Puget.
\newblock Explanation-based generalisation of failures.
\newblock In {\em ECAI}, pages 339--344, 1988.

\bibitem{gallier-jlp84}
W.~F. Dowling and J.~H. Gallier.
\newblock Linear-time algorithms for testing the satisfiability of
  propositional {H}orn formulae.
\newblock {\em J. Log. Program.}, 1(3):267--284, 1984.

\bibitem{een-sat03}
N.~E{\'{e}}n and N.~S{\"{o}}rensson.
\newblock An extensible {SAT}-solver.
\newblock In {\em SAT}, pages 502--518, 2003.
\newblock MiniSat 2.2 is available from:
  \url{https://github.com/niklasso/minisat.git}.

\bibitem{gent-jair13}
I.~Gent.
\newblock Optimal implementation of watched literals and more general
  techniques.
\newblock {\em Journal of Artificial Intelligence Research}, 48:231--252, 2013.

\bibitem{giunchiglia-ecai06}
E.~Giunchiglia and M.~Maratea.
\newblock Solving optimization problems with {DLL}.
\newblock In {\em ECAI}, pages 377--381, 2006.

\bibitem{mazure-aaai14}
{\'{E}}.~Gr{\'{e}}goire, J.~Lagniez, and B.~Mazure.
\newblock An experimentally efficient method for {(MSS, CoMSS)} partitioning.
\newblock In {\em AAAI}, pages 2666--2673, 2014.

\bibitem{hmms-aicomm15}
F.~Heras, A.~Morgado, and J.~Marques{-}Silva.
\newblock {MaxSAT}-based encodings for group {MaxSAT}.
\newblock {\em {AI} Commun.}, 28(2):195--214, 2015.

\bibitem{itai-jlp87}
A.~Itai and J.~A. Makowsky.
\newblock Unification as a complexity measure for logic programming.
\newblock {\em J. Log. Program.}, 4(2):105--117, 1987.

\bibitem{junker-aaai04}
U.~Junker.
\newblock {QuickXplain}: Preferred explanations and relaxations for
  over-constrained problems.
\newblock In {\em AAAI}, pages 167--172, 2004.

\bibitem{parsia-iswc07}
A.~Kalyanpur, B.~Parsia, M.~Horridge, and E.~Sirin.
\newblock Finding all justifications of {OWL} {DL} entailments.
\newblock In {\em ISWC}, pages 267--280, 2007.

\bibitem{kalyanpur-eswc06}
A.~Kalyanpur, B.~Parsia, E.~Sirin, and B.~C. Grau.
\newblock Repairing unsatisfiable concepts in owl ontologies.
\newblock pages 170--184, 2006.

\bibitem{nieuwenhuis-cav06}
S.~K. Lahiri, R.~Nieuwenhuis, and A.~Oliveras.
\newblock {SMT} techniques for fast predicate abstraction.
\newblock In {\em CAV}, pages 424--437, 2006.

\bibitem{liffiton-cpaior13}
M.~H. Liffiton and A.~Malik.
\newblock Enumerating infeasibility: Finding multiple {MUSes} quickly.
\newblock In {\em CPAIOR}, pages 160--175, 2013.

\bibitem{lpmms-cj15}
M.~H. Liffiton, A.~Previti, A.~Malik, and J.~Marques-Silva.
\newblock Fast, flexible {MU}s enumeration.
\newblock {\em Constraints}, 2015.
\newblock Online version:
  \url{http://link.springer.com/article/10.1007/s10601-015-9183-0}.

\bibitem{liffiton-jar08}
M.~H. Liffiton and K.~A. Sakallah.
\newblock Algorithms for computing minimal unsatisfiable subsets of
  constraints.
\newblock {\em J. Autom. Reasoning}, 40(1):1--33, 2008.

\bibitem{ludwig-ore14}
M.~Ludwig.
\newblock Just: a tool for computing justifications {w.r.t.} {ELH} ontologies.
\newblock In {\em ORE}, 2014.

\bibitem{ludwig-jelia14}
M.~Ludwig and R.~Pe{\~{n}}aloza.
\newblock Error-tolerant reasoning in the description logic $\mathcal{EL}$.
\newblock In {\em JELIA}, pages 107--121, 2014.

\bibitem{mp-tr15}
N.~Manthey and R.~Pe{\~{n}}aloza.
\newblock Exploiting {SAT} technology for axiom pinpointing.
\newblock Technical Report LTCS 15-05, Chair of Automata Theory, Institute of
  Theoretical Computer Science, Technische Universit{\"{a}}t Dresden, April
  2015.
\newblock Available from
  \url{https://ddll.inf.tu-dresden.de/web/Techreport3010}.

\bibitem{mshjpb-ijcai13}
J.~Marques{-}Silva, F.~Heras, M.~Janota, A.~Previti, and A.~Belov.
\newblock On computing minimal correction subsets.
\newblock In {\em IJCAI}, pages 615--622, 2013.

\bibitem{msjb-cav13}
J.~Marques-Silva, M.~Janota, and A.~Belov.
\newblock Minimal sets over monotone predicates in {B}oolean formulae.
\newblock In {\em CAV}, pages 592--607, 2013.

\bibitem{mpms-ijcai15}
C.~Menc\'{\i}a, A.~Previti, and J.~Marques-Silva.
\newblock Literal-based {MCS} extraction.
\newblock In {\em IJCAI}, 2015.
\newblock To Appear.

\bibitem{meyer-aaai06}
T.~A. Meyer, K.~Lee, R.~Booth, and J.~Z. Pan.
\newblock Finding maximally satisfiable terminologies for the description logic
  $\mathcal{ALC}$.
\newblock In {\em AAAI}, pages 269--274, 2006.

\bibitem{minoux-ipl88}
M.~Minoux.
\newblock {LTUR:} {A} simplified linear-time unit resolution algorithm for
  {H}orn formulae and computer implementation.
\newblock {\em Inf. Process. Lett.}, 29(1):1--12, 1988.

\bibitem{meyer-rr11}
K.~Moodley, T.~Meyer, and I.~J. Varzinczak.
\newblock Root justifications for ontology repair.
\newblock In {\em RR}, pages 275--280, 2011.

\bibitem{nguyen-dl12}
H.~H. Nguyen, N.~Alechina, and B.~Logan.
\newblock Axiom pinpointing using an assumption-based truth maintenance system.
\newblock In {\em DL}, 2012.

\bibitem{osullivan-aaai07}
B.~O'Sullivan, A.~Papadopoulos, B.~Faltings, and P.~Pu.
\newblock Representative explanations for over-constrained problems.
\newblock In {\em AAAI}, pages 323--328, 2007.

\bibitem{parsia-www05}
B.~Parsia, E.~Sirin, and A.~Kalyanpur.
\newblock Debugging {OWL} ontologies.
\newblock In {\em WWW}, pages 633--640, 2005.

\bibitem{penaloza-kr10}
R.~Pe{\~{n}}aloza and B.~Sertkaya.
\newblock On the complexity of axiom pinpointing in the {EL} family of
  description logics.
\newblock In {\em KR}, 2010.

\bibitem{pms-aaai13}
A.~Previti and J.~Marques{-}Silva.
\newblock Partial {MUS} enumeration.
\newblock In {\em AAAI}, pages 818--825, 2013.

\bibitem{rector-97}
A.~L. Rector and I.~R. Horrocks.
\newblock Experience building a large, re-usable medical ontology using a
  description logic with transitivity and concept inclusions.
\newblock In {\em Workshop on Ontological Engineering}, pages 414--418, 1997.

\bibitem{reiter-aij87}
R.~Reiter.
\newblock A theory of diagnosis from first principles.
\newblock {\em Artif. Intell.}, 32(1):57--95, 1987.

\bibitem{giunchiglia-aicomm13}
E.~D. Rosa and E.~Giunchiglia.
\newblock Combining approaches for solving satisfiability problems with
  qualitative preferences.
\newblock {\em {AI} Commun.}, 26(4):395--408, 2013.

\bibitem{schlobach-ijcai03}
S.~Schlobach and R.~Cornet.
\newblock Non-standard reasoning services for the debugging of description
  logic terminologies.
\newblock In {\em IJCAI}, pages 355--362, 2003.

\bibitem{schlobach-jar07}
S.~Schlobach, Z.~Huang, R.~Cornet, and F.~van Harmelen.
\newblock Debugging incoherent terminologies.
\newblock {\em J. Autom. Reasoning}, 39(3):317--349, 2007.

\bibitem{sebastiani-cade09}
R.~Sebastiani and M.~Vescovi.
\newblock Axiom pinpointing in lightweight description logics via {Horn-SAT}
  encoding and conflict analysis.
\newblock In {\em CADE}, pages 84--99, 2009.

\bibitem{sebastiani-tr15}
R.~Sebastiani and M.~Vescovi.
\newblock Axiom pinpointing in large $\mathcal{EL}^+$ ontologies via {SAT} and
  {SMT} techniques.
\newblock Technical Report DISI-15-010, DISI, University of Trento, Italy,
  April 2015.
\newblock Under Journal Submission. Available as
  \url{http://disi.unitn.it/~rseba/elsat/elsat_techrep.pdf}.

\bibitem{sioutos-jbi07}
N.~Sioutos, S.~de~Coronado, M.~W. Haber, F.~W. Hartel, W.~Shaiu, and L.~W.
  Wright.
\newblock {NCI} thesaurus: {A} semantic model integrating cancer-related
  clinical and molecular information.
\newblock {\em Journal of Biomedical Informatics}, 40(1):30--43, 2007.

\bibitem{parsia-jws07}
E.~Sirin, B.~Parsia, B.~C. Grau, A.~Kalyanpur, and Y.~Katz.
\newblock Pellet: {A} practical {OWL-DL} reasoner.
\newblock {\em J. Web Sem.}, 5(2):51--53, 2007.

\bibitem{slaney-ecai14}
J.~Slaney.
\newblock Set-theoretic duality: {A} fundamental feature of combinatorial
  optimisation.
\newblock In {\em ECAI}, pages 843--848, 2014.

\bibitem{spackman-amia97}
K.~A. Spackman, K.~E. Campbell, and R.~A. C{\^{o}}t{\'{e}}.
\newblock {SNOMED} {RT:} a reference terminology for health care.
\newblock In {\em AMIA}, 1997.

\bibitem{vescovi-phd11}
M.~Vescovi.
\newblock {\em Exploiting {SAT} and {SMT} Techniques for Automated Reasoning
  and Ontology Manipulation in Description Logics}.
\newblock PhD thesis, University of Trento, 2011.

\end{thebibliography}


\end{document}